%% file: paper.tex
\documentclass[a4paper]{article}

\usepackage{INTERSPEECH2021}
\usepackage{multirow}
\usepackage{todonotes}
\presetkeys{todonotes}{inline, noinlinepar}{}
\usepackage{hyperref}
\usepackage{caption}
\usepackage{subcaption}
\usepackage{diagbox}
\usepackage{rotating}
\usepackage{mathtools} 
\mathtoolsset{showonlyrefs}

\newcommand{\compare}{ComParE21~\cite{Schuller21-TI2}}

\title{Visual Transformers for Primates Classification and Covid Detection}
\name{Steffen Illium, Robert Müller, Andreas Sedlmeier and Claudia-Linnhoff Popien}
\address{Mobile and Distributed Systems Group,\\ LMU Munich}
\email{\{steffen.illium,robert.mueller,andreas.sedlmeier,linnhoff\}@ifi.lmu.de}

\begin{document}

\maketitle
\begin{abstract}
  \input{content/0_abstract}
\end{abstract}

\input{./images/data/data}

\input{content/1_introduction}

\input{content/2_dataset_task}

\input{content/3_related_work}

\input{./images/models/models}
\input{content/4_methods}

\input{./images/results/results_conf_roc}
\input{./images/results/parameter_importance}
\input{content/5_results}

\input{./images/results/all_models_uar}

\bibliographystyle{IEEEtran}
\bibliography{mybib}


\end{document}

%% file: content/0_abstract.tex
We apply the vision transformer, a deep machine learning model build around the attention mechanism, on mel-spectrogram representations of raw audio recordings. 
When adding mel-based data augmentation techniques and sample-weighting, we achieve comparable performance on both (PRS and CCS challenge) tasks of ComParE21, outperforming most single model baselines.
We further introduce overlapping vertical patching and evaluate the influence of parameter configurations.
 

%% file: images/data/data.tex
\begin{figure*}[ht]
     \centering
     \hfill
     \begin{subfigure}[b]{0.25\textwidth}
         \centering
         \includegraphics[width=\textwidth]{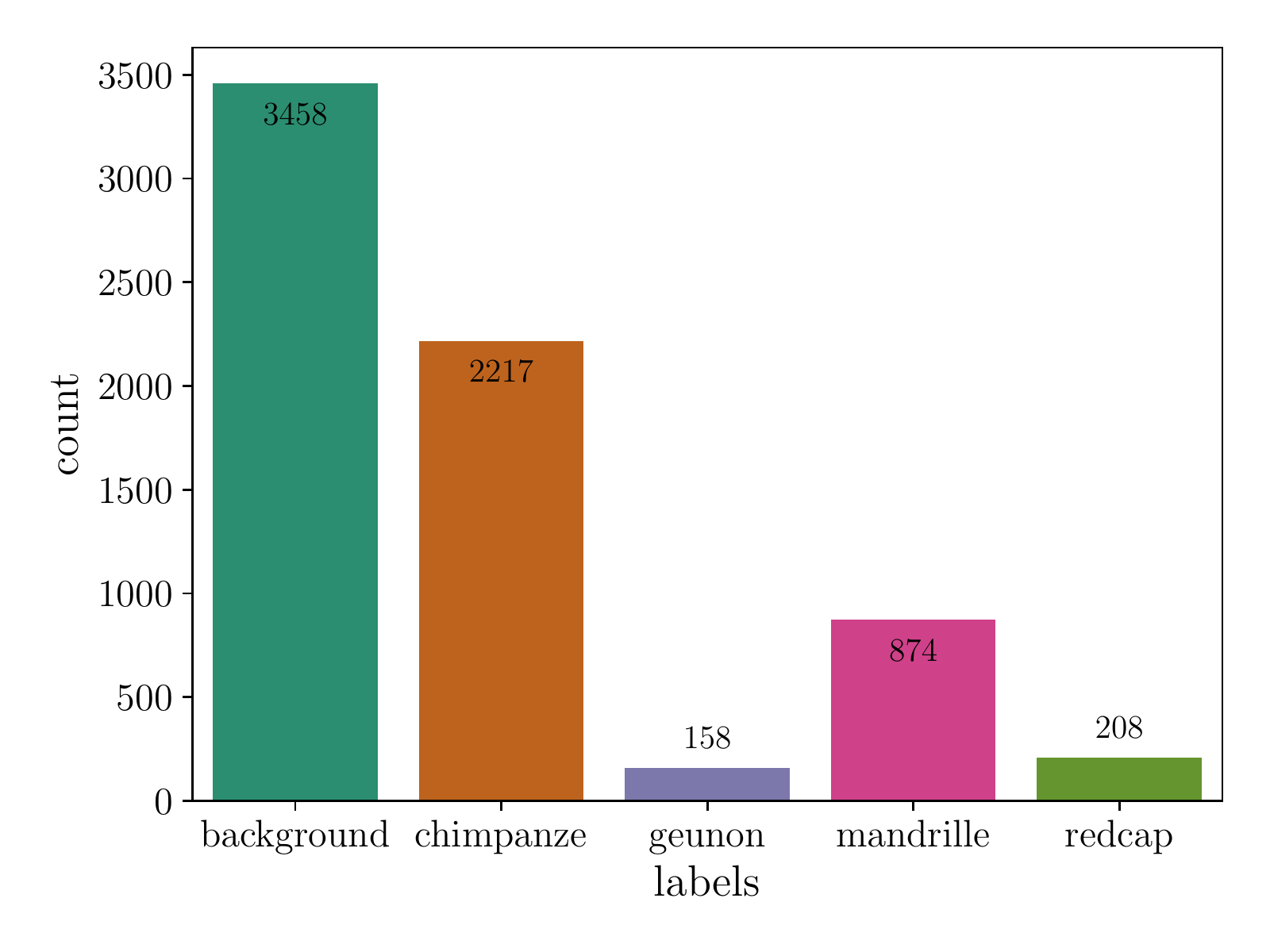}
         \caption{PRS-train: class counts}
         \label{fig:primates_train_counts}
     \end{subfigure}
     \hfill
     \begin{subfigure}[b]{0.25\textwidth}
         \centering
         \includegraphics[width=\textwidth]{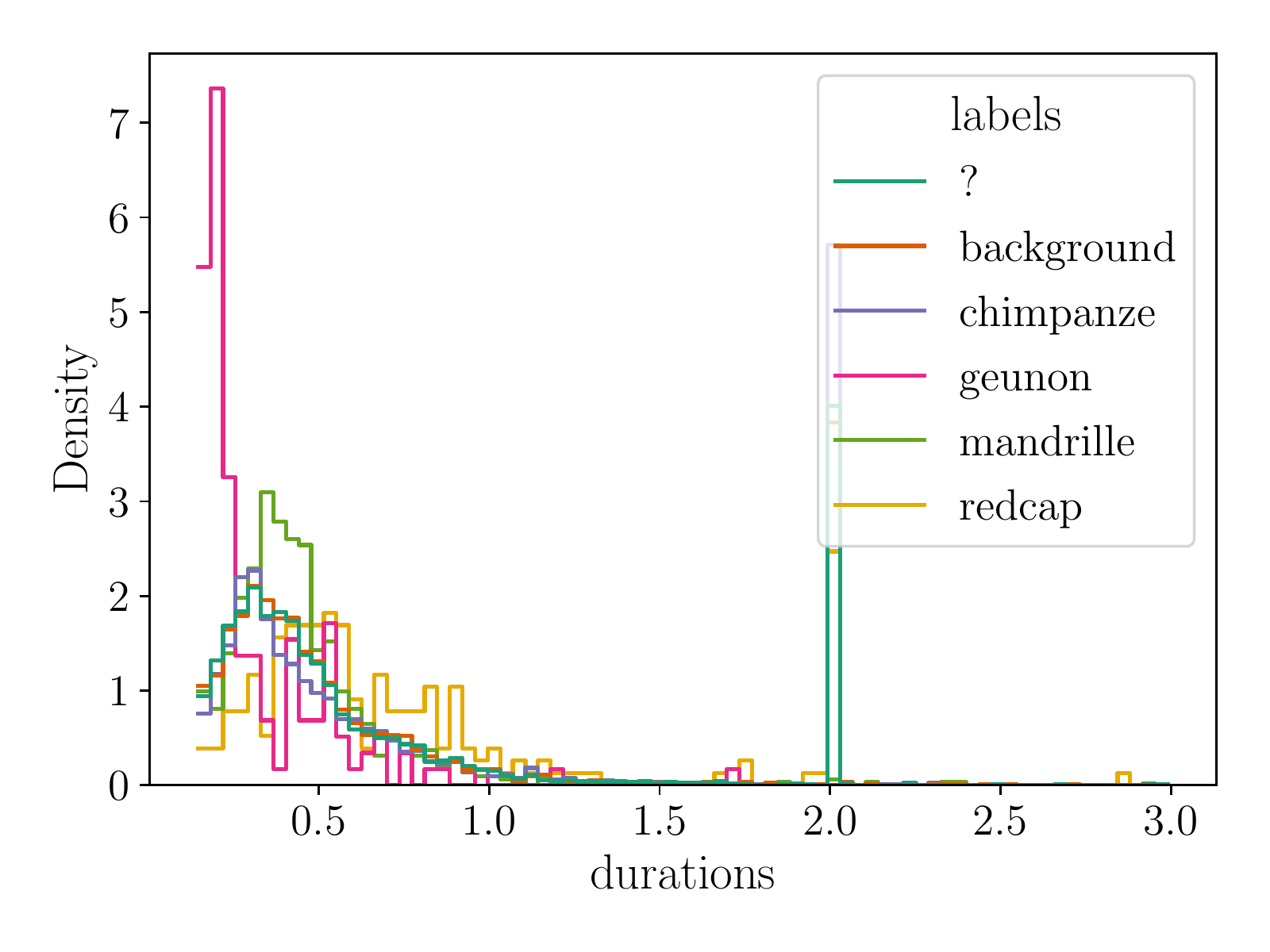}
         \caption{PRS-train: Durations per class}
         \label{fig:primates_train_durations}
     \end{subfigure}
     \hfill
     \begin{subfigure}[b]{0.25\textwidth}
        \centering
        \includegraphics[width=\textwidth]{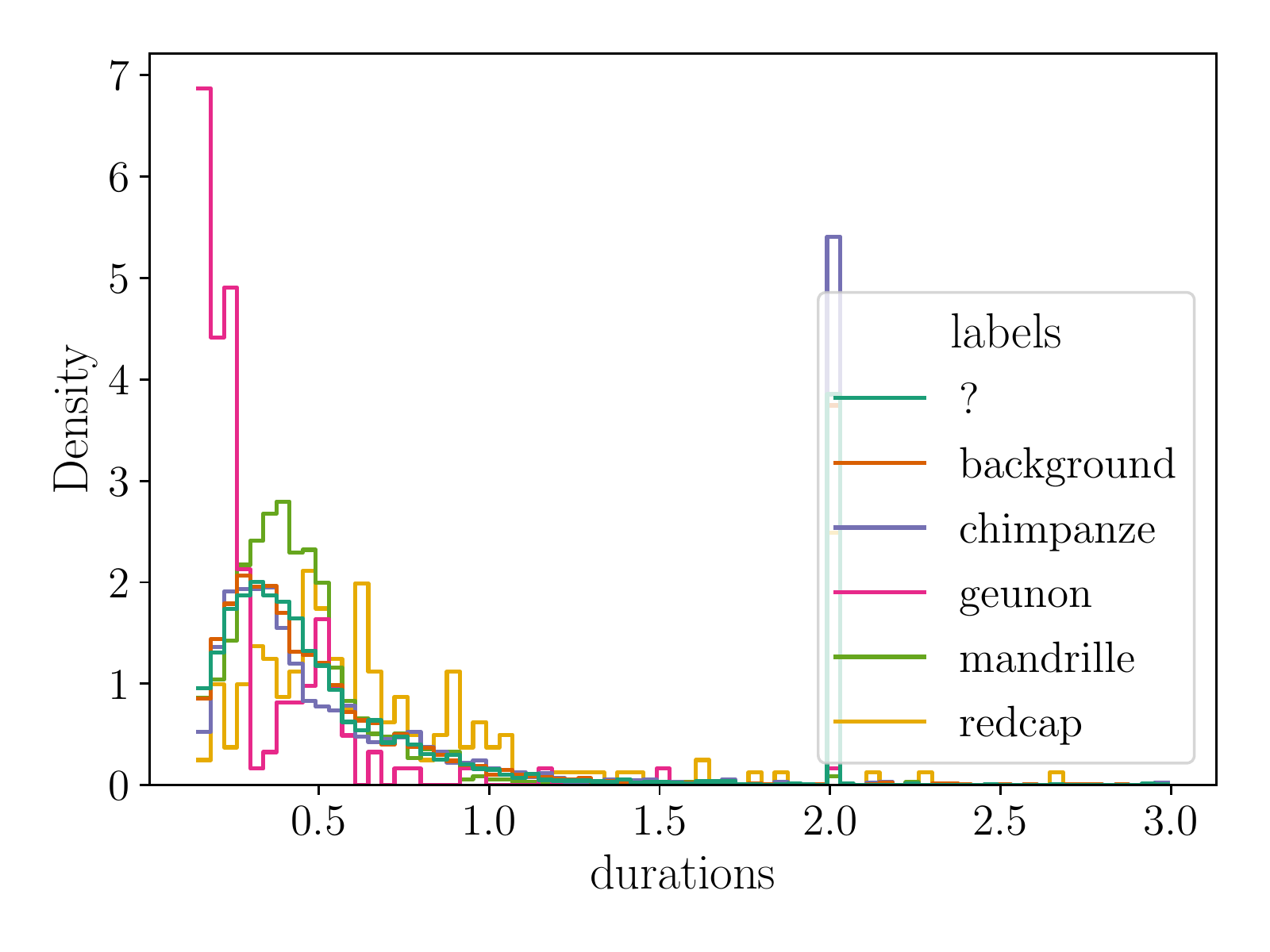}
         \caption{PRS-devel: Durations per class}
         \label{fig:primates_devel_durations}
     \end{subfigure}
    \hfill
    ~\\
    \centering
    \hfill
    \begin{subfigure}[b]{0.25\textwidth}
         \centering
         \includegraphics[width=\textwidth]{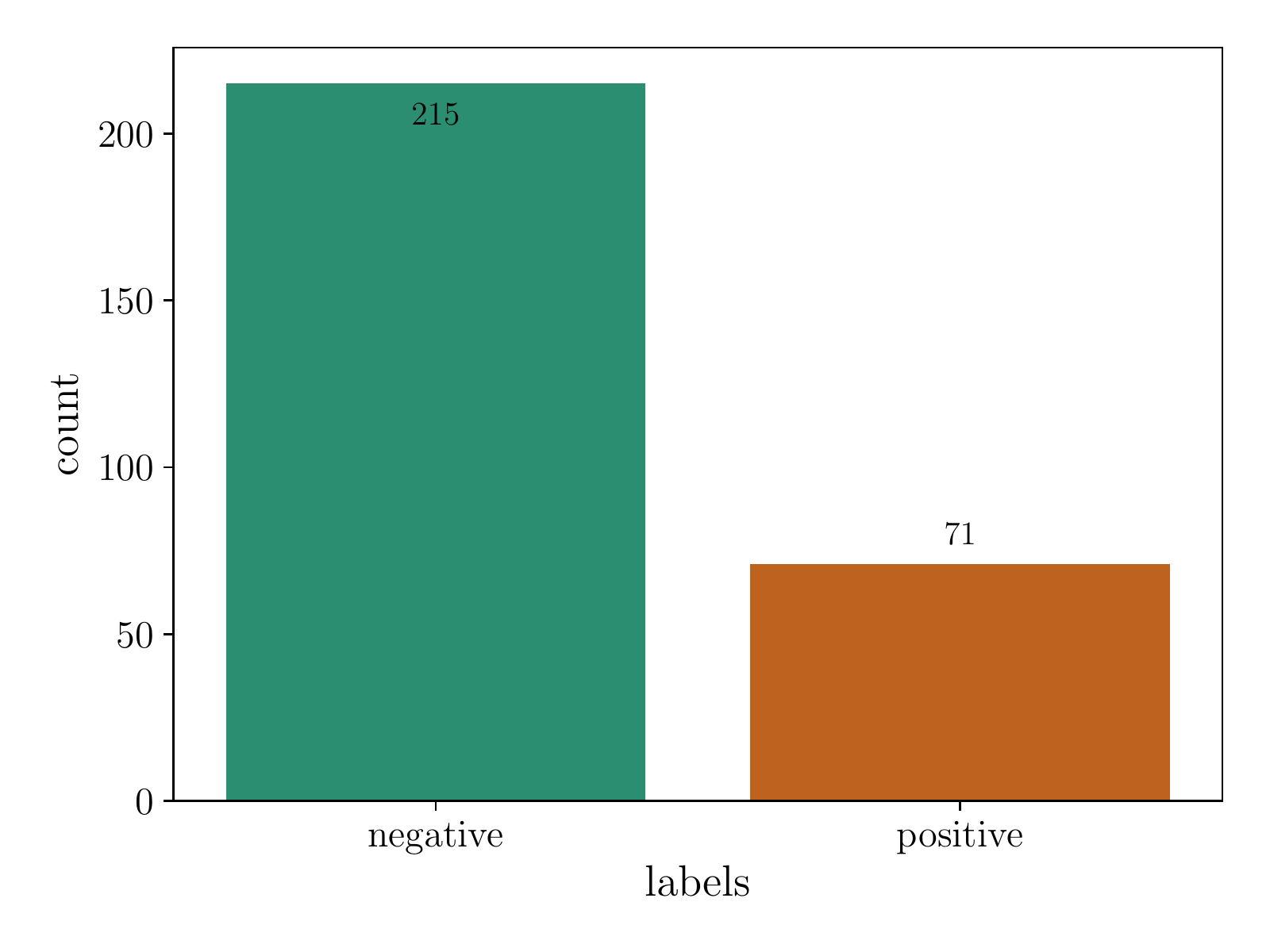}
         \caption{CCS-train: class counts}
         \label{fig:ccs_train_counts}
    \end{subfigure}
    \hfill
    \begin{subfigure}[b]{0.25\textwidth}
         \centering
         \includegraphics[width=\textwidth]{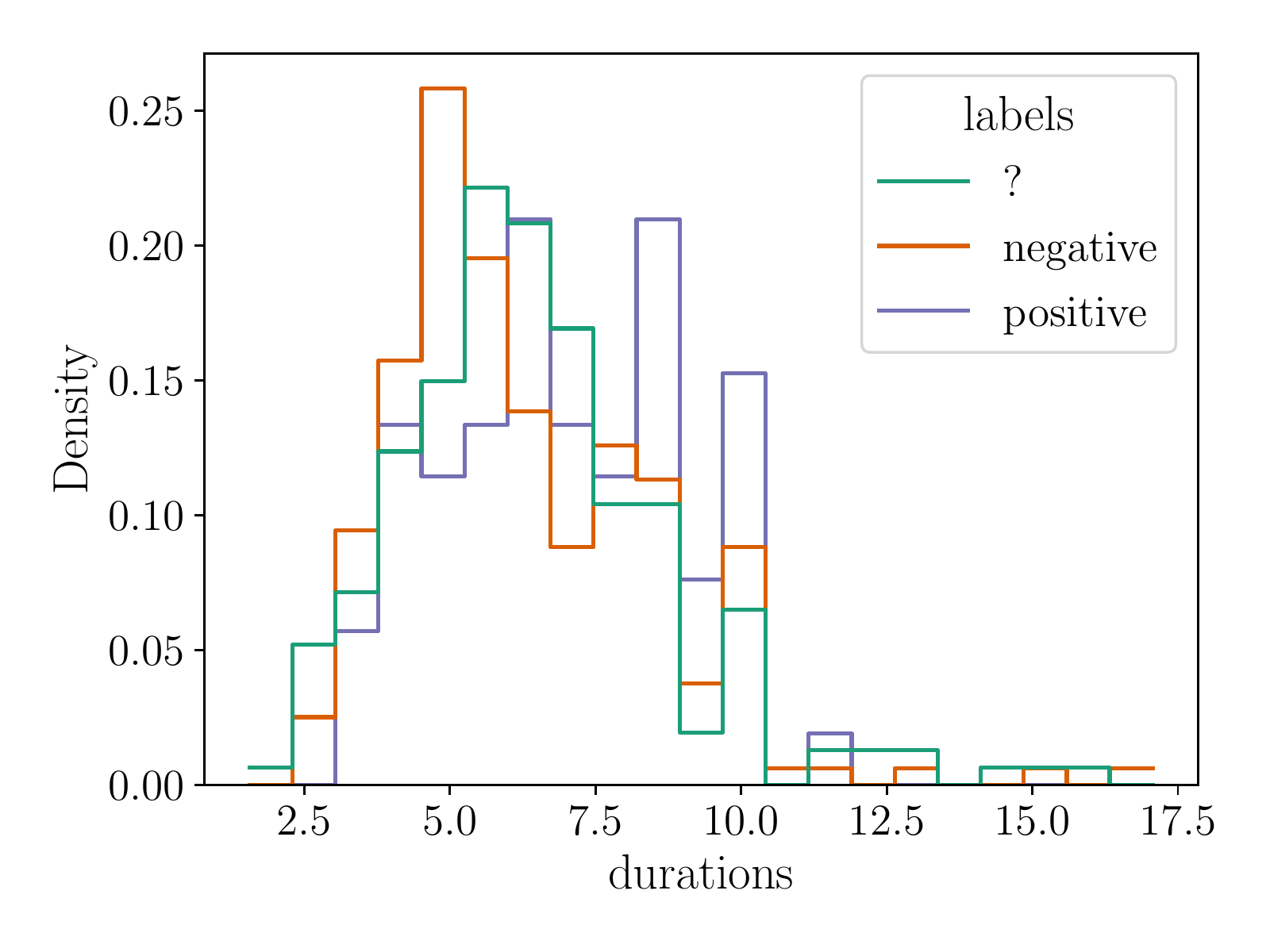}
         \caption{CCS-train: Durations per class}
         \label{fig:ccs_train_durations}
     \end{subfigure}
     \hfill
     \begin{subfigure}[b]{0.25\textwidth}
         \centering
         \includegraphics[width=\textwidth]{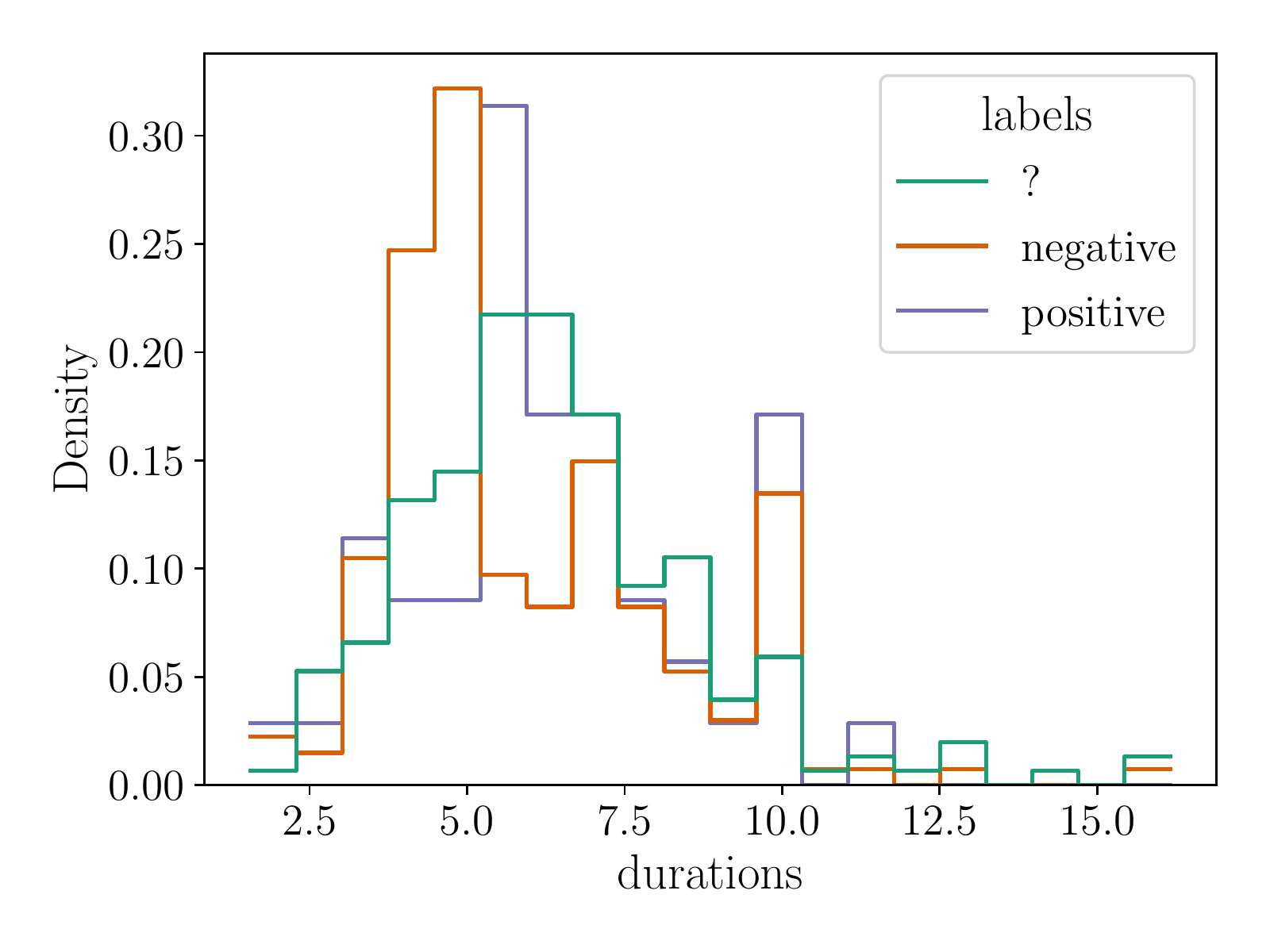}
         \caption{CCS-devel: Durations per class}
         \label{fig:ccs_devel_durations}
     \end{subfigure}
     \hfill
     \hfill
     \caption{Statistics for 'primates' and 'CCS' datasets. \ref{fig:primates_train_counts} and \ref{fig:ccs_train_counts} show the counts per class for each training set, \ref{fig:primates_train_durations}, \ref{fig:primates_devel_durations} \ref{fig:ccs_train_durations} \&  \ref{fig:ccs_devel_durations} show the distribution of file-durations per class. For comparison, the test set duration (label: ?) is also shown.}
     \label{fig:data}
\end{figure*}

%% file: content/1_introduction.tex
\section{Introduction}
\label{sec:introduction}

As raw audio can meaningful be represented  by mel-spectrograms, the field of audio classification has seen a lot of influences from the field of computer vision. 
Such algorithm work mostly out of the box (once a suitable hyper-parameter combination is found) and build a competitive baseline for many problems. 
Currently the most used architecture from the range of deep-learning algorithms is the convolutional neural network (CNN) \cite{fukushima1982neocognitron}.
When additional written text is provided alongside raw audio, algorithms from the field of natural language processing (NLP) take over.
Resulting embeddings from both types of algorithms can subsequently be combined before entering a joint classifier.
Most recently, transformers, employing deep-learning attention mechanism, are taking over the field of NLP. 
The VisionTransformer~(ViT) ~\cite{dosovitskiy2020visiontransformer} recently fused both of the fields by interpreting images as sequences of patches and thus utilising what can be called dynamic convolution. 
We relate to this idea by training the ViT on mel-spectrogram representations (thus single channel images) of raw audio data, which is a novelty.
We further observe the performance impact by varying the shape and order of patches to match the natural temporal dependency in spectrograms. 
Additionally, we incorporated data-augmentation techniques while over-sampling the datasets to overcome limitations in two of the challenge datasets of \compare.

%% file: content/2_dataset_task.tex
\section{Dataset and Task Description}
\label{sec:data}

We focus on two of the provided datasets of \compare, one binary and one multi-class classification task:

\textbf{Primates Sub-Challenge (PRS)}:
Zwerts et al.~\cite{zwerts2021introducing_primates}~describes the Primate Vocalisations Corpus, which consists of non-invasive acoustic recordings from Central Africa Mefou Primate Sanctuary to identify and count species. This is a multi-class classification task including the classes: ['\textit{background}', '\textit{chimpanze}', '\textit{geunon}', '\textit{mandrille}', '\textit{redcap}'], which are either primate species or background noises from the natural primate habitat. The \textit{background} class is ambiguous as there is the possibility of faint primate scream mixing.
For further introductory explanation please refer to~\cite{Schuller21-TI2, zwerts2021introducing_primates}.
When working with the dataset, we noticed an imbalance across the classes, in the train-~(\autoref{fig:primates_train_counts})~as well as in the provided \textit{devel}-dataset. 
The counts per sample and class are similar in \textit{train} \& \textit{devel} (c.p.~\autoref{fig:primates_train_counts}), but we noticed slight variations in audio-sample length (number of frames, c.p.~\autoref{fig:primates_train_durations}~\&~\ref{fig:primates_devel_durations}). 
Distributions for \textit{train} and \textit{devel} are comparable while the test dataset has variations in sample length regarding the number of very small ($<= 0.3$~seconds). Class \textit{genuon} e.g., not only is underrepresented but also comes at small audio sample lengths.

\textbf{The COVID-19 Cough Sub-Challenge (CCS)}:
Brown et al.~\cite{brown2020exploring_CCS} used an application (Android and Web) to gather (crowdsource) coughs and breathing from \textit{negative} and \textit{positive} tested humans~\textit{COVID-19}~(binary classification task).
For further introductory explanation please refer to~\cite{Schuller21-TI2, brown2020exploring_CCS}.
This binary dataset not only is imbalanced, also the number of samples is rather small (215\textit{p} vs. 71\textit{n}, c.p.~\autoref{fig:ccs_train_counts}), especially in respect to usual deep learning datasets. 
Please notice that samples are five times as long as compared to PRS-challenge (up to 16 minutes).
When comparing the sample length distribution (c.p.~\autoref{fig:ccs_train_durations}~\&~\ref{fig:ccs_devel_durations}) we notice slight differences in comparison of \textit{devel} and \textit{test}.

Scores are measured by the \textit{unweighted average recall}~(UAR, also known as the \textit{Balanced Error Rate}~(BER)~\cite{rosenberg2012classifying}):
\vspace{-0.3em}
\begin{equation}
    \vspace{-0.3em}
    UAR = 0.5 \times \bigg(\frac{TP}{FN+TP} + \frac{TN}{TN+FP}\bigg)
    \label{equ:uar_ber}
\end{equation}

%% file: content/3_related_work.tex
\section{Related Work}
\label{sec:related_work}

\subsection{Audio Classification}
Research in the field of audio classification can be categorized based on the respective goal which is to be achieved, e.g., genre prediction \cite{dieleman2014end} or classification \cite{piczak2015environmental} (multi-class classification approaches), as well as binary classification or anomaly detection.

As is the case for most research fields concerned with classification tasks, progress in recent years was mostly achieved using deep neural networks (DNN) and gradient based learning techniques.
By contrast to earlier research, features are no longer manually constructed (as was the case e.g., for work based on mel-frequency cepstral coefficients (MFCC)) but instead automatically inferred (e.g. VGG \cite{simonyan2014very}).
Consequently, these approaches can be differentiated based on how the input data is structured:
i) direct usage of the mel-spectrograms, i.e., the time-frequency representation or ii) end-to-end learning from the audio's wave-forms.

When using spectrogram features, it is possible to apply and build upon network architectures that were conceived to tackle problems related to image data.
Research has shown that these approaches outperform classical methods relying on manually constructed features \cite{piczak2015environmental, hershey2016cnn}.
While \cite{dieleman2014end} has shown that it is possible to construct end-to-end learning approaches directly using the audio's wave-form, their method's performance is worse compared to spectrogram-based approaches.
\cite{dai2016deep} was finally able to match the performance of CNN based architectures using log-mel spectrograms by applying very deep networks.

Only recently, it has been shown that transformer architectures \cite{vaswani2017attention} can be a competitive approach to CNNs in the image domain. Initially developed in natural language processing (NLP), the transformer architecture aimed to improve on complex sequence models like recurrent convolutional neural networks. 
Still, the transformer follows the encoder-decoder structure as is common for state-of-the art sequence models based around the attention mechanism by mapping a sequence of input features $x$ to a sequence of representations $z$.
The so-called Vision Transformer (ViT)~\cite{dosovitskiy2020visiontransformer} showed that the architecture is able to achieve competitive results to convolutional neural networks by interpreting an image as a sequence of patches.

In the work at hand, we analyze the performance of such a ViT architecture when applied to audio data by using the spectrogram representation of the data as input features as well as the importance of its hyper-parameters.

\subsection{Audio Data Augmentation}

To improve results in audio classification tasks, data augmentation techniques are commonly applied.
One such approach is masking or \textit{SpecAugmentation}~\cite{specaug}, which masks areas (which can be restricted by a maximum size) of certain frequency bands and temporal bins by replacing them with random shapes of zeros.
When combined with time warping, research has shown that these techniques can reduce over-fitting in training.
Results in \cite{illium2020surgical} show that these and other approaches like loudness modifications, time shifting, or the addition of noise improve classification results of CNN based models.
In the work at hand, we build upon these results by implementing and evaluating augmentation techniques on the given dataset of the \compare.



%% file: images/models/models.tex
\begin{figure*}[htb]
     \centering
     \begin{subfigure}[b]{0.27\textwidth}
         \centering
         \includegraphics[width=\textwidth]{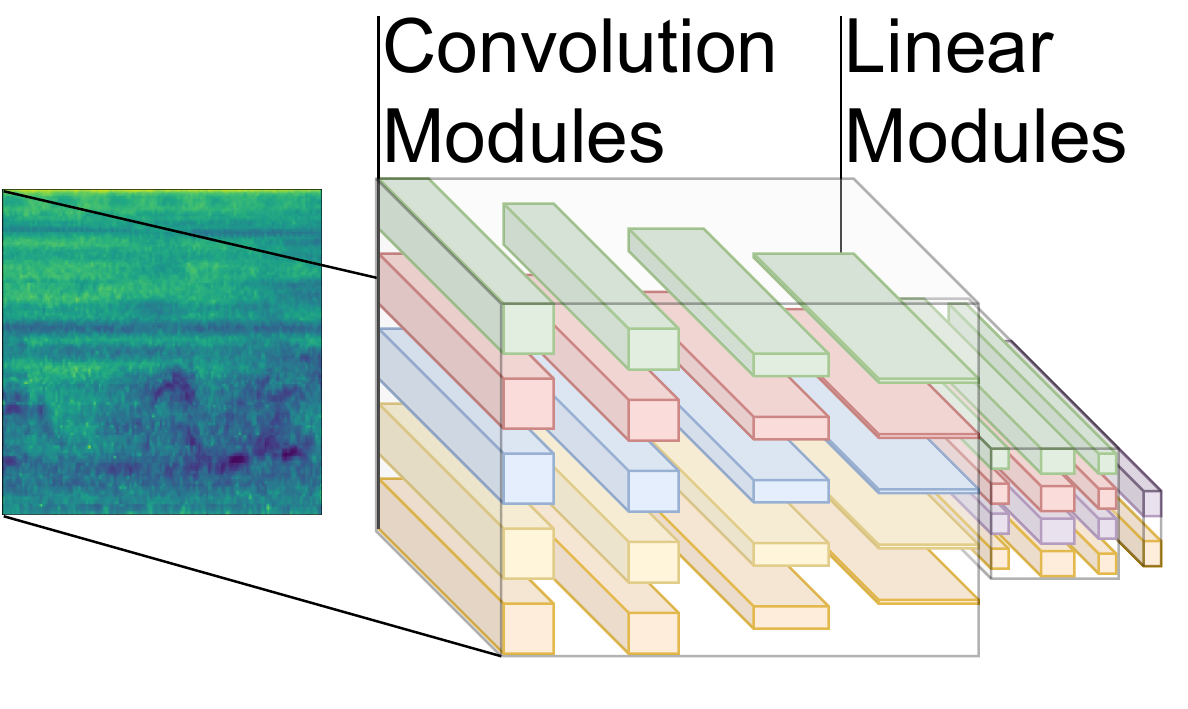}
         \caption{Convolutional Baseline (CNN)}
         \label{fig:cnn}
     \end{subfigure}
     \hfill
     \begin{subfigure}[b]{0.27\textwidth}
         \centering
         \includegraphics[width=\textwidth]{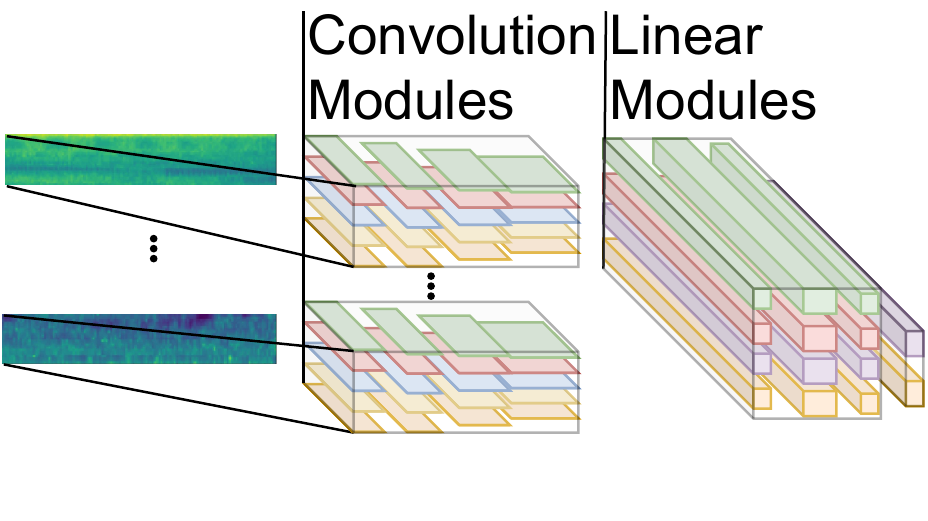}
         \caption{SubSpectral Classifier (SSC)}
         \label{fig:ssc_model}
     \end{subfigure}
     \hfill
     \begin{subfigure}[b]{0.27\textwidth}
         \centering
         \includegraphics[width=\textwidth]{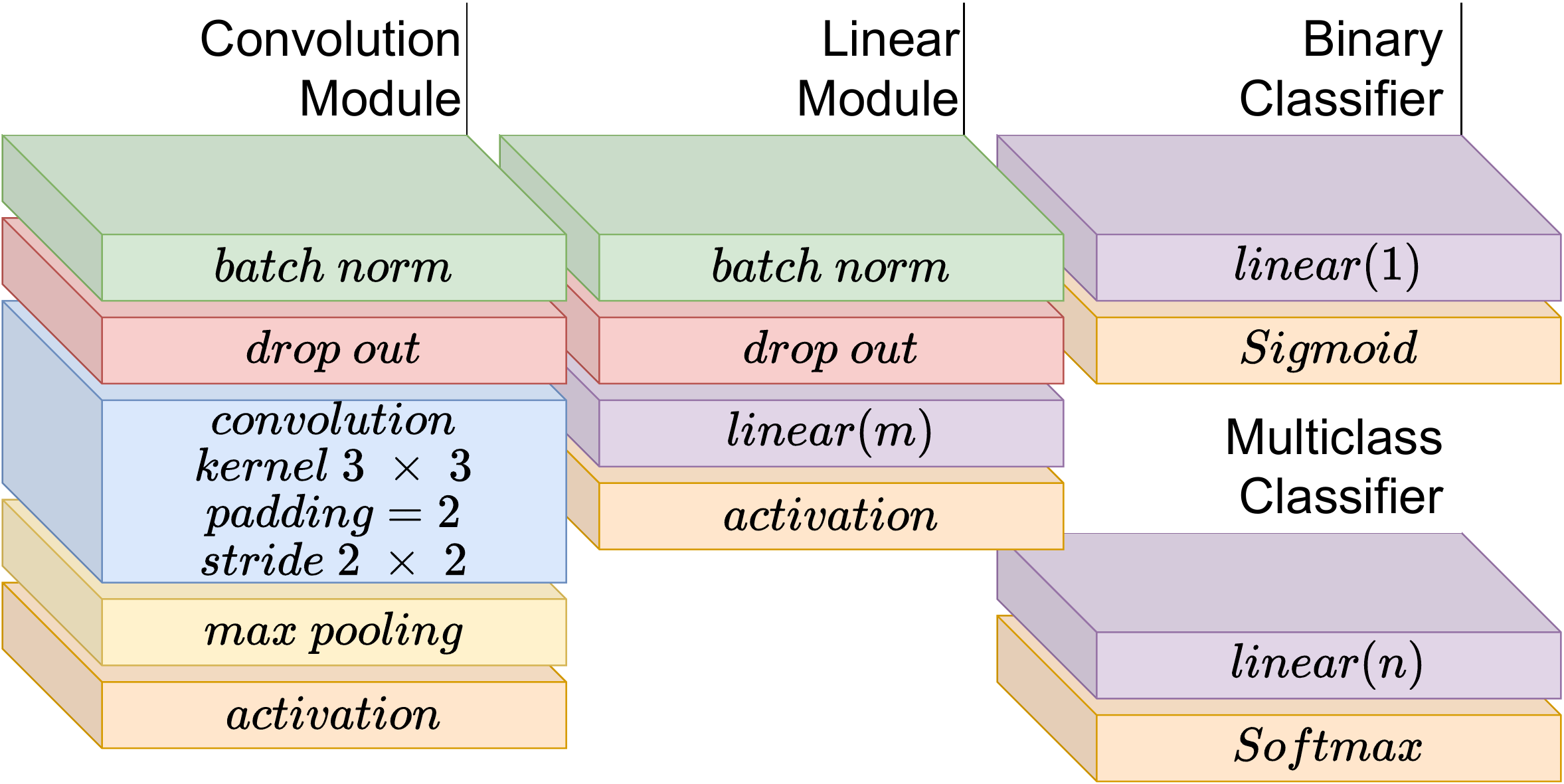}
         \caption{CNN Module description}
         \label{fig:cnn_blocks}
     \end{subfigure}
    ~\\
    \begin{subfigure}[b]{0.27\textwidth}
         \centering
         \includegraphics[width=\textwidth]{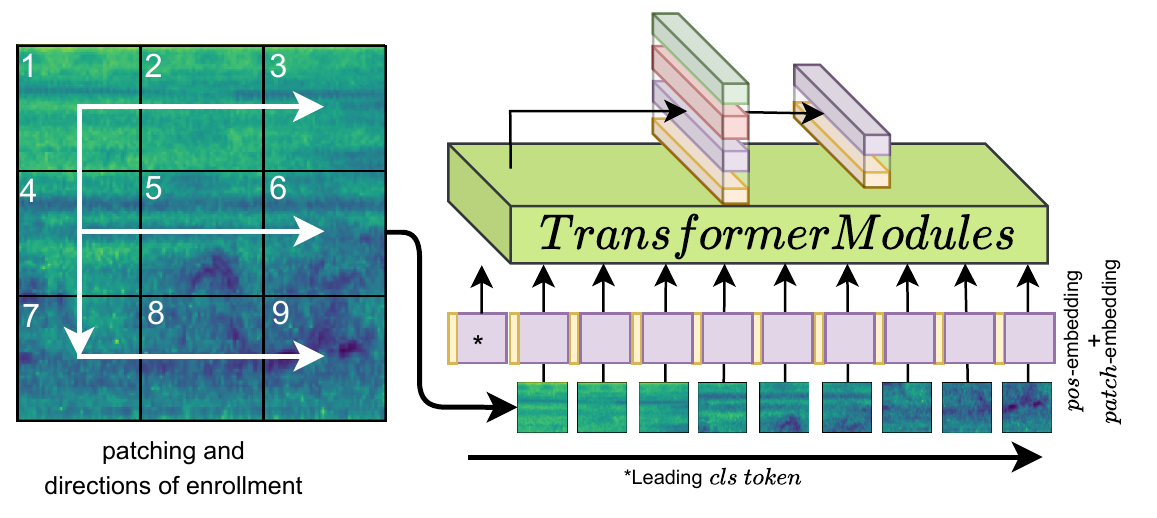}
         \caption{VisionTransformer (ViT) architecture \& application}
         \label{fig:vit_model}
    \end{subfigure}
    \hfill
    \begin{subfigure}[b]{0.27\textwidth}
         \centering
         \includegraphics[width=\textwidth]{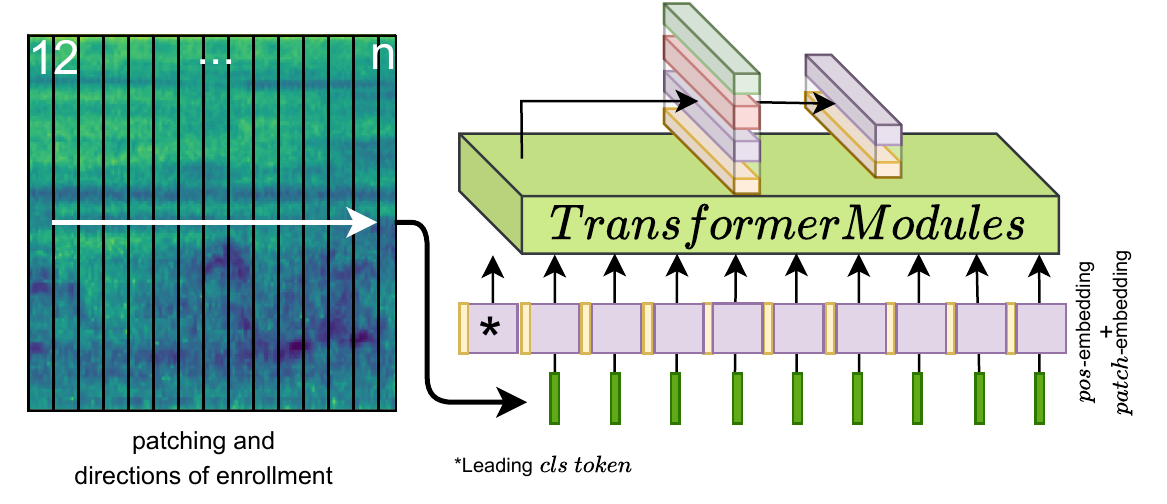}
         \caption{VerticalVisionTransformer (ViT) architecture \& application}
         \label{fig:vvit_model}
     \end{subfigure}
     \hfill
     \begin{subfigure}[b]{0.27\textwidth}
         \centering
         \includegraphics[angle=0, width=\textwidth]{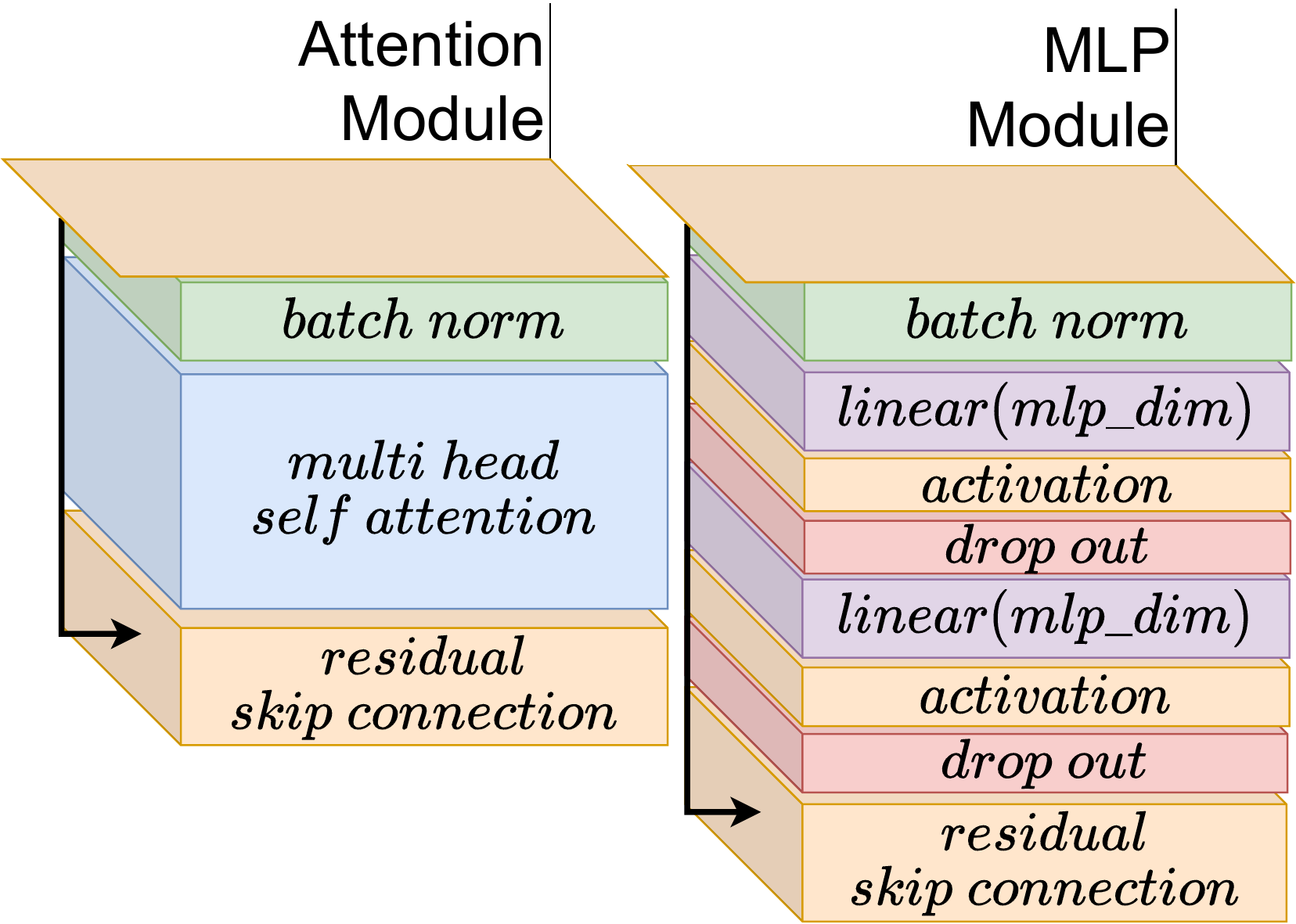}
         \caption{Transformer component description}
         \label{fig:attn_desc}
     \end{subfigure}
     \hfill
        \caption{Model architectures as introduced and described in \autoref{sec:methods}. Hyper-parameters such as number of neurons or layer depth are determined per model and dataset in a broad hyper-parameter search.}
        \label{fig:models}
\vspace{-1em}
\end{figure*}

%% file: content/4_methods.tex
\section{Methods}
\label{sec:methods}


Based on the observations from previous section, we decide to apply data augmentation as presented in~\cite{illium2020surgical} while over-sampling the datasets to re-balance the deficits in data distribution per class and length.
Oversampling was assumed to work best at equally distributed classes, so that the total number of samples ($S_{tot}$) $S_{tot}=max(len(S_{class})) \times n\_classes$, we then sampled randomly with replacements given the probability $p = 1 / len(S_{class})$ for each sample and class respectively.
This is a critical point. 
Without equal sampling, no training was possible for both challenges with the algorithm introduced in Subsection~\ref{sec:model_architectures}. 
The audio samples were further cropped to smaller length. We let the HP-search find the best matching sample length for each dataset and algorithm. 
This results in shorter sample length $(\sim~0.4)$ for PRS and longer samples $(\sim~1.2)$ for CCS dataset .
This chapter now briefly revisits the data-augmentation techniques, then introducing the deep learning models.

\subsection{Mel-Spectrogram Augmentation}
\label{sec:data_augmentation}
Named parameters in this section are treated as model hyper-parameter and further tuned by the tree-structured Parzen Estimator-algorithm (TPE) hyper-parameter search~\cite{akiba2019optuna}.
We searched a meaningful parameter-space in $400 \text{runs}$ per algorithm ($200 \text{training-epochs}$) and dataset. 
For all augmentation every draw from a random distribution is performed per sample draw.

~

~
\textbf{Shift Augmentation:}~
Temporal offset is provided, to account for positional variations within the the mel-spectrograms. 
Data-samples are shifted either left or right ($direction \sim Bernoulli(0.5)$) by $shift \sim \mathcal{U}(0, shift\text\_ratio)$ percent of axis length in the temporal dimension. Resulting zero-value data-points $s$ are left empty.

\indent\textbf{Noise Augmentation:}~
Random Gaussian noise, $noise \sim \mathcal{N}(0, noise\text\_ratio)$, is added ($S+S\times noise$) to the mel-spectrogram to improve robustness and generalisation capabilities of end-to-end trained neural networks models. 
This method has already been successfully applied in tasks of audio classification (where recordings can be noise) and image processing.

\indent\textbf{SpecAugment:}~
As presented by \cite{ren2019spec} we randomly \textit{mask} vertical and horizontal windows (mel- and temporal-axis) with $zero$ value (all data-points $s in w$).
First a random starting point, $s \sim U(0, T)$, is determined. Then we draw the size of the window by $w \sim \mathcal{U}(0, mask\text\_ratio)$. 

\indent\textbf{Loudness Augmentation:}~
We adjust the loudness (intensity) of recording by a loudness factor. $l\sim \mathcal{U}(0, loudness\_ratio)$ determines how much of the original signal is added to the original sample ($S + S \times l$). Loud signals get even louder instead just raising all values (which would be normalised anyway).

\subsection{Model Architectures}
\label{sec:model_architectures}
In our experiments we trained and applied three different architectures in an end to end fashion, by \textit{AdamW}~\cite{loshchilov2018decoupled_adamw} as optimiser for gradient based back-propagation.
For comparison, we choose a similar parameter range where possible settings for all the models, if not specified otherwise. Please note the varying size and introduction of additional parameters through changes in network architecture. 
We further tuned the model hyper-parameters as described above by TPE~\cite{akiba2019optuna}. 

For both tasks (PRS and CCS) we define default parameters as follows, PRS: $\text{n\_logits}=5$, $\text{loss}=ce\_loss$, CCS: $\text{n\_logits}=1$, $\text{loss}=bce\_loss$. Further model parameters are the same to reduce the influence of a wide hyper-parameter space. Those are $\text{dropout}=0.2$, \textit{GELU}~\cite{hendrycks2016gaussianerrorlinearunits}

\indent\textbf{CNNBaseline~(CNN)}~
As deep-learning baseline, we implement a quite regular convolutional neural network~(CNN~\autoref{fig:cnn}). 
Four blocks of batch-normalisation~\cite{ioffe2015batch}, dropout, convolution max-pooling and a activation function are stacked. 
Filter sizes are $\{32, 64, 128, 64\}$ while convolutional-kernel sizes are $3\times3$. with $\text{zero-padding}=1$ we kept the shape of each activation matrix. 
Pooling kernel sizes are $2\times2$, cutting the resolution in half, finally. 
The last convolution is followed by a linear module composed of batch-normalisation, dropout, fully-connected layer (128 neurons) and an activation function. The final fully connected classifier is implemented as described above and dependent on the task.

\indent\textbf{SubSpectralClassifier~(SSC)}~
We consider another deep-learning baseline as comparison for the ViT algorithms. The SubSpectralClassifier~(SSC) is implemented as proposed by ~\cite{illium2020surgical}. In short; four small convolutional neural networks are trained on different non-overlapping of mel-bands $n\_mels=128/8$, depicted in \autoref{fig:ssc_model}.
Those concatenated band-wise embeddings are then processed by a classifier sub-network (three fully-connected layers with [128, 64, n] neurons, respectively). 
We switched the ReLU activation function in favour of the more recent GELU activation function.
Further parameters are the same as in~\cite{illium2020surgical}, this includes batch-normalisation and position and rate of dropout.

\indent\textbf{VisionTransformer~(ViT)}~
In reference to~\cite{dosovitskiy2020visiontransformer} we implement the ViT as stack of blocks of multi-head self-attention (scaled dot-product attention) followed by double layered mlps (GELU activated fully-connected layers with dropout, c.p.~\autoref{fig:vit_model}~\&~\ref{fig:attn_desc}), both wrapped with residual skip-connections~\cite{he2016deep_resnet}. 
Multi-head attention runs multiple self-attention operations in parallel, then subsequently fusing the individual embeddings into a single $d/n\text{\_}heads$ sized embedding.
This style of architecture was first seen in form of the \textit{Transformer} neural- network~\cite{vaswani2017attention_transformer} which relates a single sentence (sequences of word embedding) to it self.

\vspace{-1.5em}
\begin{align}
    \text{Attention}(Q, K, V) &= softmax(\frac{QK^T}{\sqrt{n}})V~\\ \label{eq:attention}
    \text{MultiHead}(Q, K, V &= [\text{head}_1;...;\text{head}_n]W^O\\ \label{eq:multi_head}
    \text{where head}_i &= \text{Attention}(QW_i^Q,KW_i^K,VW_i^V)
\end{align}
An additional linear module is attached before evaluating the resulting embedding by the subsequent classifier~(c.p.~\autoref{fig:cnn_blocks}).

\indent\textbf{VerticalVisionTransformer (VViT)}~
We strictly follow the ViT design described above but realise overlapping full-height (mel-dimension) patches to analyse the mel-spectrogram in an more intuitive fashion. 
Since mel-spectrograms posses a temporal relation, we move the kernel ($n \text\_ mels \times patch \text\_ size, stride=1$) horizontally along the temporal axis rather then in a rows-then-column pattern~(c.p.~\autoref{fig:vit_model}~vs.~\ref{fig:vvit_model}).
The width of the overlapping patches was empirically found to work best between 5 and 9 pixels. This procedure results in more and larger patches (in comparison to the original ViT) while respecting the natural order of the audio/mel-spectrogram domain.

%% file: images/results/results_conf_roc.tex
\begin{figure}[htb]
     \centering
     \hfill
     \begin{subfigure}[b]{0.23\textwidth}
        \centering
        \includegraphics[width=\textwidth]{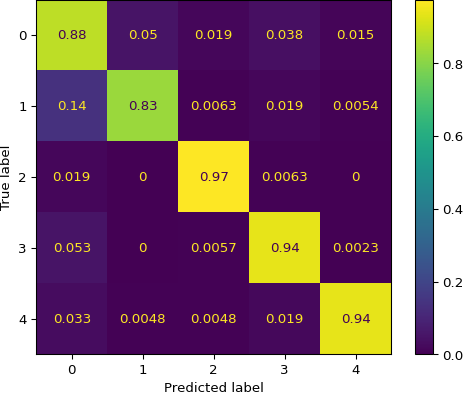}
        \caption{Confusion matrix}
        \label{fig:confusion_matrix_VT_7889c_ViT_primates}
     \end{subfigure}
     \hfill
     \begin{subfigure}[b]{0.23\textwidth}
        \centering
        \includegraphics[width=\textwidth]{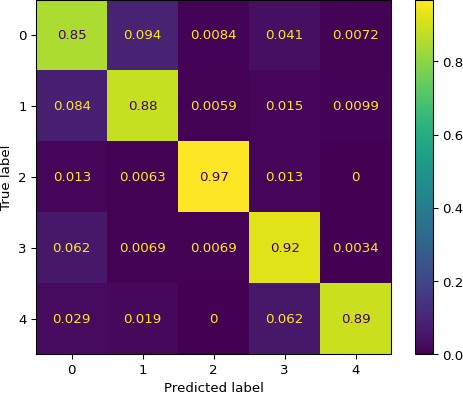}
        \caption{Confusion matrix}
        \label{fig:confusion_matrix_VVT_c054665_VViT_primates}
     \end{subfigure}
     \hfill
     ~\\
     \hfill
     \begin{subfigure}[b]{0.23\textwidth}
          \centering
         \includegraphics[width=\textwidth]{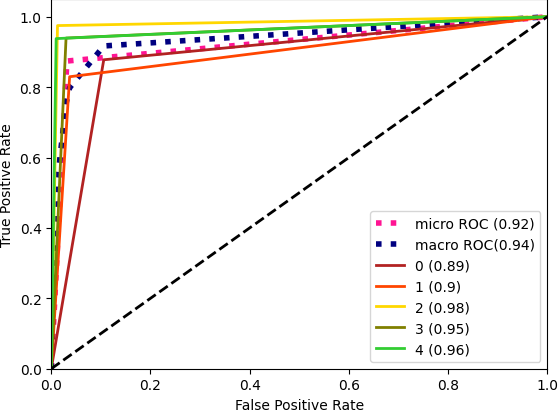}
         \caption{ROC plot}
     \label{fig:roc_plots_VT_7889c_ViT_primates}
     \end{subfigure}
     \hfill
     \begin{subfigure}[b]{0.23\textwidth}
          \centering
         \includegraphics[width=\textwidth]{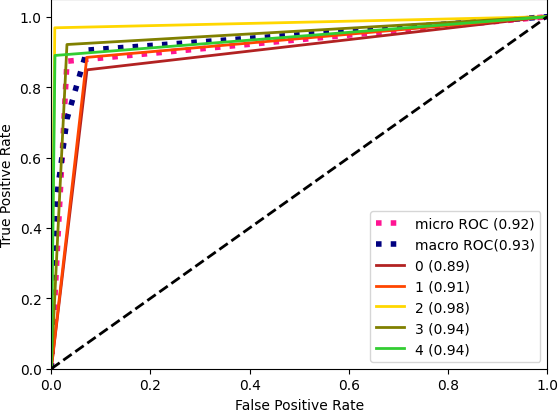}
         \caption{ROC plot}
     \label{fig:roc_plots_VVT_c054665_VViT_primates}
     \end{subfigure}
     \hfill
     \caption{Additional Evaluation Metrics for ViT~(\ref{fig:confusion_matrix_VT_7889c_ViT_primates},\ref{fig:roc_plots_VT_7889c_ViT_primates}) \& VViT~(\ref{fig:confusion_matrix_VVT_c054665_VViT_primates},\ref{fig:roc_plots_VVT_c054665_VViT_primates}) on PRS-dataset. For confusion matrices, rows sum up to 1. Class assignments are: \{\textbf{0}: background, \textbf{1}: chimpanze, \textbf{2}: geunon, \textbf{3}: mandrille, \textbf{4}: redcap\}}
     \label{fig:additional_evaluation_metrics_roc_conf}
     \vspace{-1.6em}
\end{figure}

%% file: images/results/parameter_importance.tex
\begin{figure}[htb]
     \centering
     \hfill
     \begin{subfigure}[b]{0.23\textwidth}
        \centering
        \includegraphics[width=\textwidth]{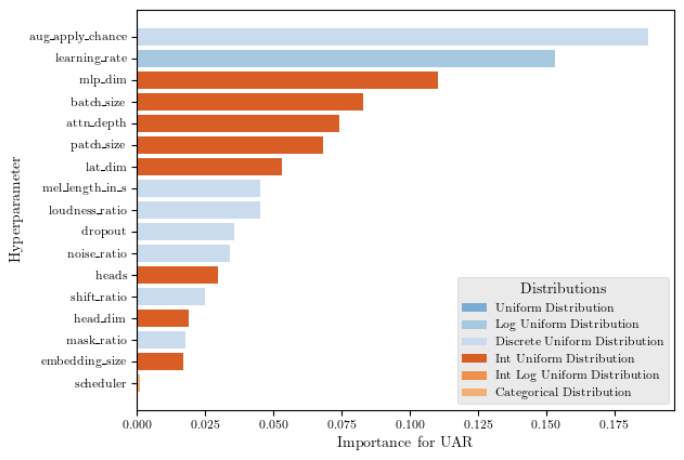}
        \caption{ViT parameter importance}
        \label{fig:vit_hp_importance}
     \end{subfigure}
     \hfill
     \begin{subfigure}[b]{0.23\textwidth}
          \centering
         \includegraphics[width=\textwidth]{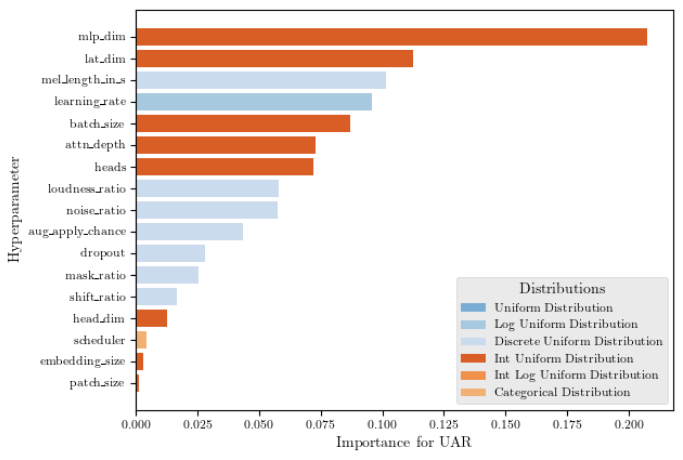}
         \caption{VViT parameter importance}
     \label{fig:vvit_hp_importance}
     \end{subfigure}
     \hfill
     \caption{ViT~(\ref{fig:vit_hp_importance})~\&~VViT~(\ref{fig:vvit_hp_importance}) hyper-parameter importance on PRS dataset measured by \textit{fANOVA}\cite{pmlr-v32-hutter14_fANOVA}}
     \label{fig:paramter_importance}
     \vspace{-1.2em}
\end{figure}

%% file: content/5_results.tex
\section{Results}
\label{sec:results}

We train the described algorithms for 200 epochs on the over-sampled and augmented \textit{train} datasets.
No sample from \textit{devel} or \textit{test} is touched in training while seeds $(0, ... ,19)$ are fixed along models and runs. 
Hyper-parameters are tuned by TPE as stated above per network architecture (CNN,~SSC,~ViT~\&~VViT)~(c.p.\autoref{sec:methods}).

Evaluations based on \textit{devel} show, that most of the provided baselines of this years classification tasks of \textit{ComParE 2021} are mostly out-performed by even the simplest CNN architectures (CNN~\&~SSC), with \textit{OpenXBOW} and \textit{AUDeep} beeing the strongest competitors. 
Depending on the task, the \textit{Transformer}-style approaches (ViT~\&~VVit) then overtake both (ComParE and our CNN baselines) by a margin (Figure \ref{fig:model_performance_boxplot}), outperforming the \textit{devel} dataset performance measured in UAR (c.p., Eq. \ref{equ:uar_ber}). Model performance on \textit{test} seemed to stagnate in most cases. 

For PRS we observe high rate of confusion between \textit{background}~(0) and \textit{chimpanze}~(1) on both attention based models~(c.p.~\autoref{fig:additional_evaluation_metrics_roc_conf}). 
We noticed the same behaviour for CNN~\&~SSC.
Presumably more chimpanzees are mixed faintly into the background class. 
We therefore suggest further steps of data-processing.
For CCS we noticed a high false negative rate (up to $60\%$) along all models, which is very bad in respect to a usage of medical applicability.
Usually a high specificity over a high sensitivity is required.

We further analysed the hyper-parameter importance for both attention based models (ViT~\&~VViT, c.p.~\autoref{fig:paramter_importance}). While both models do not care whether a learning rate scheduler (\textit{lr\_scheduler}~in [None, $\text{\textit{lr}}=n^{epoch}, n \sim \mathcal{U}(0.88, 1)$] is present, the initial given \textit{lr} and total available parameters of fully connected layers (\textit{mlp\_dim}~\&~\textit{lat\_dim}) are important for both models.
Usually transformer models are known to work best, employing a specific training sequence, including warm-up-phases and sophisticated schedulers. In both tasks, we could not confirm these practices.

On one hand, we are surprised by the low importance of both \textit{embedding\_size}~\&~\textit{head\_dim} for both models, as these parameters control how much of information can be evaluated at every single layer in parallel (the lower the initial projection dimension, the higher the initial information compression rate). 
On the other hand, we find it interesting to see how the importance of data augmentation not only depends on the used datasets, but on the model architecture as well. Even when they differ very little.

We further observe a reduced overall performance of VViT (at lower variance) which boils down to the assumption, that spatial, non overlapping regions are more important for both tasks, than a bigger picture, which processes information on all mel bands at once.

In future we plan to extend our view to more recent models featuring attention, e.g. the \textit{Performer}\cite{choromanski2020rethinking_performer} which approximates the result of the regular attention mechanism, without explicitly constructing the quadratic-sized attention matrix.

%% file: images/results/all_models_uar.tex
\input{tables/results}
\begin{figure}
    \centering
    \includegraphics[width=0.75\columnwidth]{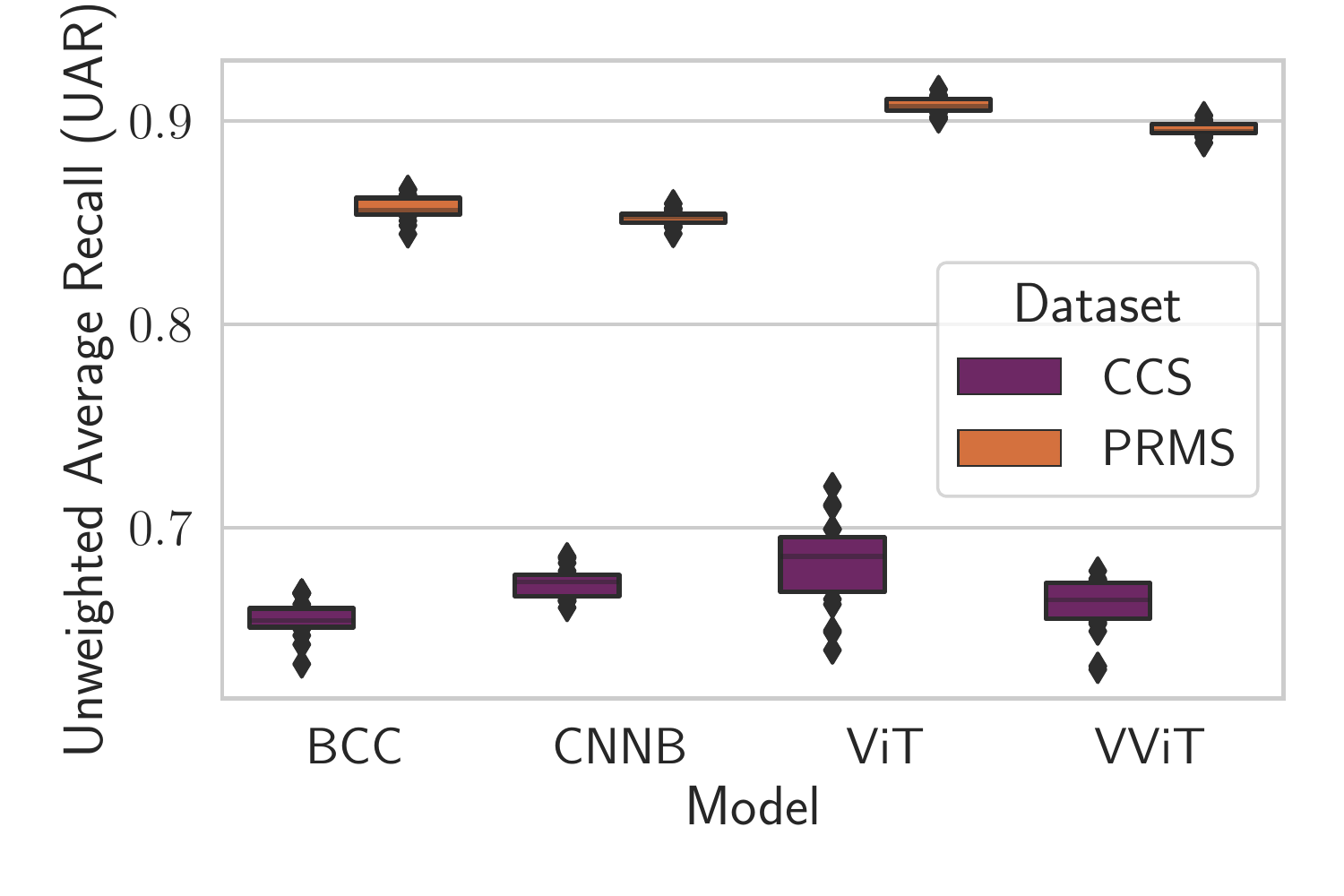}
    \caption{Model performance measured in UAR over 20 seeds}
    \label{fig:model_performance_boxplot}
    \vspace{-1.5em}
\end{figure}

%% file: tables/results.tex
\begin{table}[hbt]
\begin{tabular}{r|cc|c|c}
\textbf{Dataset}                             & \multicolumn{2}{c|}{\textbf{primates}}      & \multicolumn{2}{c}{\textbf{coughing}}                     \\ \hline
\multicolumn{1}{l|}{\textbf{Model}}              & \multicolumn{1}{c|}{Devel} & Test & Devel                 & \multicolumn{1}{c}{Test} \\ \hline
\multicolumn{1}{l|}{Deep Spectrum} & \multicolumn{1}{c|}{81.3}  & 78.8 & 63.3 & \multicolumn{1}{c}{64.1} \\ \hline
\multicolumn{1}{l|}{End2You}       & \multicolumn{1}{c|}{72.70} & 70.8 & 61.8 & \multicolumn{1}{c}{64.7} \\ \hline
\multicolumn{1}{l|}{OpenSMILE}     & \multicolumn{1}{c|}{82.4}  & 82.2 & 61.4 & \multicolumn{1}{c}{65.5} \\ \hline
\multicolumn{1}{l|}{OpenXBOW}      & \multicolumn{1}{c|}{83.3}  & 83.9 & 64.7 & \multicolumn{1}{c}{\textbf{72.9}} \\ \hline
\multicolumn{1}{l|}{AuDeep}        & \multicolumn{1}{c|}{84.6}  & 86.1 & 67.6 & \multicolumn{1}{c}{67.6} \\ \hline
\multicolumn{1}{l|}{\textbf{Fusion}}        & \multicolumn{1}{c|}{-}     & 87.5 & - & \multicolumn{1}{c}{73.9} \\ \hline\hline
\multicolumn{1}{l|}{ours CNN}      & \multicolumn{1}{c|}{85.9}      &  -    & \multicolumn{1}{l|}{68.6} & \multicolumn{1}{l}{67.4}     \\ \hline
\multicolumn{1}{l|}{ours SSC}      & \multicolumn{1}{c|}{86.6}      &   84.2   & \multicolumn{1}{l|}{66.8} & \multicolumn{1}{l}{72.0}     \\ \hline
\multicolumn{1}{l|}{ours ViT}      & \multicolumn{1}{c|}{\textbf{91.5}}      &   \textbf{88.3}  & \multicolumn{1}{l|}{\textbf{72.0}} & \multicolumn{1}{l}{69.9}     \\ \hline
\multicolumn{1}{l|}{ours VViT}      & \multicolumn{1}{c|}{90.3}      &   87.2  & \multicolumn{1}{l|}{67.8} & \multicolumn{1}{l}{68.6}     \\ \hline
\end{tabular}
\caption{Our experimental results compared to provided baselines from \compare.} 
\label{tab:results}
\vspace{-2em}
\end{table}